\documentclass[useAMS,usenatbib,usegraphicx]{mn2e}
\usepackage{bm}
\usepackage{times}

\newcommand{\nH}{n_{\mathrm{H}}}

\title[Shattering as a production source of very small grains]
{Shattering by turbulence as a production source of very small grains}
\author[Hirashita]{Hiroyuki Hirashita\thanks{E-mail:
    hirashita@asiaa.sinica.edu.tw} 
\\
Institute of Astronomy and Astrophysics, Academia Sinica,
P.O. Box 23-141, Taipei 10617, Taiwan
}
\date{2010 June 10}

\pagerange{\pageref{firstpage}--\pageref{lastpage}} \pubyear{2010}

\begin{document}
\label{firstpage}
\maketitle

\begin{abstract}
The origin of grain size distribution in the interstellar
medium is one of the most fundamental problems in
the interstellar physics. In the Milky Way, smaller
grains are more abundant in number, but their origins are
not necessarily specified and quantified. One of the most
efficient drivers of small grain production is interstellar
turbulence, in which dust grains can acquire
relative velocities large enough to be
shattered. Applying the framework of shattering
developed in previous papers, we show that
small ($a\la 0.01~\micron$) grains reach the
abundance level observed in the Milky Way
in $\sim 10^8$ yr (i.e.\ within the grain lifetime)
{by shattering in warm neutral medium}.
We also show that if part of grains experience
{additional}
shattering in warm ionized medium,
carbonaceous grains with
$a\sim 0.01~\micron$ are redistributed into
smaller sizes. This could explain the relative
enhancement of
very small carbonaceous grains with
$a\sim 3$--100 \AA. Our theory also explains
the ubiquitous association between
large grains and very small grains naturally.
Some tests for our theory are proposed in terms
of the metallicity dependence.
\end{abstract}

\begin{keywords}
dust, extinction --- galaxies: evolution  ---
galaxies: ISM --- 
turbulence
\end{keywords}

\section{Introduction}

Formation and evolution of dust grains is one of the
fundamental problems in the interstellar physics.
In particular, the grain size distribution largely
affects the opacity (or extinction) of interstellar
medium (ISM). \citet[hereafter MRN]{mathis77}
reproduced the extinction curve of the Milky Way
by a mixture of silicate and graphite with a size
distribution of
$n(a)\propto a^{-3.5}$, where $n(a)\,\mathrm{d}a$
is the number density of grains with radii between
$a$ and $a+\mathrm{d}a$. More recently,
\citet{weingartner01} and \citet{li01}
have derived more
detailed grain size distribution, finding that
an enhancement of small carbonaceous grains,
especially polycyclic aromatic hydrocarbons (PAHs),
is required to reproduce the mid-infrared (MIR)
spectrum in the Milky Way.

The grain size distribution reflects the physical
processes in grain formation and evolution
\citep{odonnell97}. Dust grains are not only
produced and ejected by supernovae (SNe)
\citep[e.g.][]{kozasa89} and asymptotic giant
branch (AGB) stars \citep[e.g.][]{gail84} but also
processed in the ISM.
\citet[hereafter HY09]{hirashita09} show that
shattering occurs efficiently in diffuse ISM, where
grains are accelerated by magnetohydrodynamic (MHD)
turbulence. In particular, turbulence in warm
ionized medium (WIM)
can drive shattering of dust grains in
$\la 10$ Myr, producing small
grains efficiently \citep{hirashita10}. Therefore,
shattering should play a significant role in
determining the grain size distribution.
Shattering also occurs efficiently in
supernova shocks \citep*{borkowski95,jones96}.

There are some indications that the grains formed
and supplied to the ISM are biased to large sizes.
As \citet{nozawa07} show, the grains ejected from
SNe II are biased to large sizes because they
suffer from the destruction in the shocked region
before being ejected into the ISM \citep[cf.][]{bianchi07}.
The grain radius is typically around
$a\sim 0.1~\micron$. The typical size of grains
condensed in AGB stars is also suggested to be large
($a\sim 0.1~\micron$) from the observations of
spectral energy distributions
\citep{groenewegen97,gauger99},
although \citet{hofmann01} show that the grains are
not single-sized.
Moreover, in the Milky Way and other systems whose
metallicity is around solar, the major part of the
dust mass comes from the grain growth by accretion
of heavy elements in
interstellar clouds \citep[e.g.][]{dwek98,draine09}.
Not only the grain growth by accretion but also
coagulation in dense clouds causes a strong
depletion of small grains (HY09). To summarize, the
grains supplied from molecular clouds or stars
should be biased to the largest size range,
$a\sim 0.1~\micron$.

If the major part of dust grains supplied are large,
the origin of small grains is worth investigating
seriously. Here we focus on shattering as a
production source of small grains. Specifically,
we examine if large grains with $a\sim 0.1~\micron$
are efficiently redistributed into small
($\la 0.01~\micron$) grains by shattering driven
by interstellar turbulence, which is ubiquitous in the
ISM. \citet{hirashita10} have shown
that shattering in the WIM produced by starburst activities
can efficiently produce small grains
from large grains supplied from SNe II. In this
Letter, we discuss shattering in a more general
context by showing
that turbulence plays a fundamental role
in supplying small grains in more quiescent
environments {such as in warm neutral medium (WNM)
covering a large fraction of the ISM in the Milky Way.}
Shattering in supernova shocks should also be
considered \citep{jones96}, but the inclusion of
this process into our framework is left for future work.
Throughout this paper, grains are assumed to be
spherical, and the words, `large' and `small', are
used for grain radii of
$a\sim 0.1~\micron$ ($\sim$ the largest size range
in MRN) and $a\la 0.01~\micron$, respectively.

\section{Method}\label{sec:method}

\subsection{Initial condition}

{
To examine interstellar shattering induced by
turbulence as a mechanism
of producing small grains from large grains, we
assume the initial grain size
distribution to be dominated by large grains.
This `initial' grain size distribution
effectively represents not only one at the formation
in stellar ejecta but also one after the grain growth
in cold clouds in the ISM.}
The functional
form of the initial grain size distribution is represented
by a log-normal distribution,
$n(a)=n_\mathrm{ini}(a)$:
\begin{eqnarray}
n_\mathrm{ini}(a) &
\hspace*{-3mm}= & \hspace*{-3mm}  \left\{
\begin{array}{ll}
{\displaystyle\frac{C}{a}}\exp\left\{ -
{\displaystyle\frac{[\ln (a/a_0)]^2}{2\sigma^2}}
\right\}  &
(a_\mathrm{min}\leq a\leq a_\mathrm{max}) \\
0 & (\mbox{otherwise}),
\end{array}
\right. \nonumber \\
\end{eqnarray}
where $C$ is the normalizing constant determined by
equation (\ref{eq:norm}), $a_0$ and $\sigma$ are the
central grain radius and the standard deviation of the
log-normal distribution, respectively, and
$a_\mathrm{min}$ and $a_\mathrm{max}$ are the
minimum and the maximum
grain sizes, respectively. We adopt $a_0=0.1~\micron$
according to the above argument in Introduction,
$a_\mathrm{min}=3.5$ \AA,
$a_\mathrm{max}=1~\micron$, and $\sigma =0.6$.
The selection of
$\sigma$ is based on the size dispersion of large
Si and C grains as calculated by \citet{nozawa07}.

We consider two dust species, silicate and graphite,
and distribute each species according to the above
size distribution. The normalizing constant $C$ is
determined for each species by
\begin{eqnarray}
\mathcal{R}m_\mathrm{H}n_\mathrm{H}=
\int_0^\infty\frac{4\pi}{3}a^3\rho_\mathrm{gr}
n_\mathrm{ini}(a)\,\mathrm{d}a,\label{eq:norm}
\end{eqnarray}
where $\mathcal{R}$ is the dust-to-hydrogen mass
ratio (i.e.\ dust abundance relative to hydrogen),
$n_\mathrm{H}$ is the hydrogen number density, and
$\rho_\mathrm{gr}$ is the grain material density
(3.3 g cm$^{-3}$ and 2.2 g cm$^{-3}$ for silicate
and graphite, respectively). We adopt
$\mathcal{R}=4.0\times 10^{-3}$ and
$3.4\times 10^{-3}$ for silicate and graphite,
respectively, according to the typical Galactic
dust-to-gas ratio (HY09).

\subsection{Shattering}\label{subsec:shatter}

The evolution of grain size distribution by
shattering is calculated based on HY09, whose
formulation is taken from \citet{jones94,jones96}.
The shattering equation is calculated for silicate
and graphite separately to avoid the complexity caused
by the
collision between different species. The grain--grain
collision rate is estimated based on the grain velocity
as a function of grain size (Section \ref{subsec:vel}),
and if the relative velocity is
higher than the shattering threshold (2.7 and
1.2 km s$^{-1}$ for silicate and graphite,
respectively), the grains are
redistributed into smaller grains according to the
size distribution of shattered fragments, which is
assumed to be power-law ($\propto a^{-3.3}$;
note that the results are not very sensitive to the
exponent; \citealt{jones96,hirashita10}).


{
HY09 consider shattering in various ISM phases. Among
them, we first consider WNM, which occupies a
significant fraction of the interstellar space.
We also treat additional shattering in WIM. Although
WIM occupies a small fraction in the ISM, the short
time-scale of shattering in WIM (HY09) can make a
significant imprint on the grain size distribution.
}

\subsection{Grain velocities}\label{subsec:vel}

The grain velocity as a function of grain radius $a$
is taken from \citet{yan04}, who considered the grain
acceleration by hydrodrag and gyroresonance
(resonance which occurs when the Doppler-shifted
frequency of the MHD wave in the grain's guiding centre
rest frame is a multiple of the gyrofrequency) based
on \citet{yan03} and calculated the grain velocities
achieved in various ISM phases. They adopted the
following physical conditions: in WNM
$\nH =0.3$ cm$^{-3}$, $T=6000$ K,
$n_\mathrm{e}=0.03$ cm$^{-3}$, $G_\mathrm{UV}=1$,
$V_\mathrm{A}=20$ km s$^{-1}$, $L=100$ pc
($T$ is the gas temperature, $n_\mathrm{e}$ is the
electron number density, $G_\mathrm{UV}$
is the UV radiation field normalized to the typical
Galactic value in the solar neighbourhood,
$V_\mathrm{A}$ is the Alfv\'{e}n speed, and $L$ is
the injection scale
for turbulence), and in WIM $\nH =0.1$ cm$^{-3}$,
$T=8000$ K,
$n_\mathrm{e}=0.0991$, and the same values as
in WNM for the other quantities.

Large ($a>0.1~\micron$) grains are mainly accelerated
by gyroresonance. Thus, in considering shattering of
large grains, gyroresonance is the key process.
Although the grain charge strongly depends on gas
density, electron fraction, and
UV flux, the velocities of large grains do not
strongly depend on the grain charge as mentioned
in \citet{hirashita09}: The acceleration by
gyroresonance increases with the grain charge, but the
acceleration duration (the hydrodrag time-scale)
decreases with the grain charge \citep{yan03}.
Thus, as for the small grain production from large
($a\ga 0.1~\micron$) grains, the results
in this paper is insensitive to the grain charge.
As shown in \citet{hirashita10}, the velocities
of large grains do not sensitively depend on the
gas density. The grain velocity may depend on the
electron fraction, since the damping of turbulent
motion depends on it. The dependence on the ionization
fraction is roughly bracketed by the velocities in
WNM and WIM. Since the velocities of large grains are
similar in these two phases, they are not affected by
the electron fraction.
The remaining uncertainty in the model comes from
the magnetic field (or the Alfv\'{e}n velocity), which
is discussed in Section \ref{subsec:wnm}.

\section{RESULTS}\label{sec:result}

\subsection{Shattering in WNM}\label{subsec:wnm}

\begin{figure*}
\includegraphics[width=0.45\textwidth]{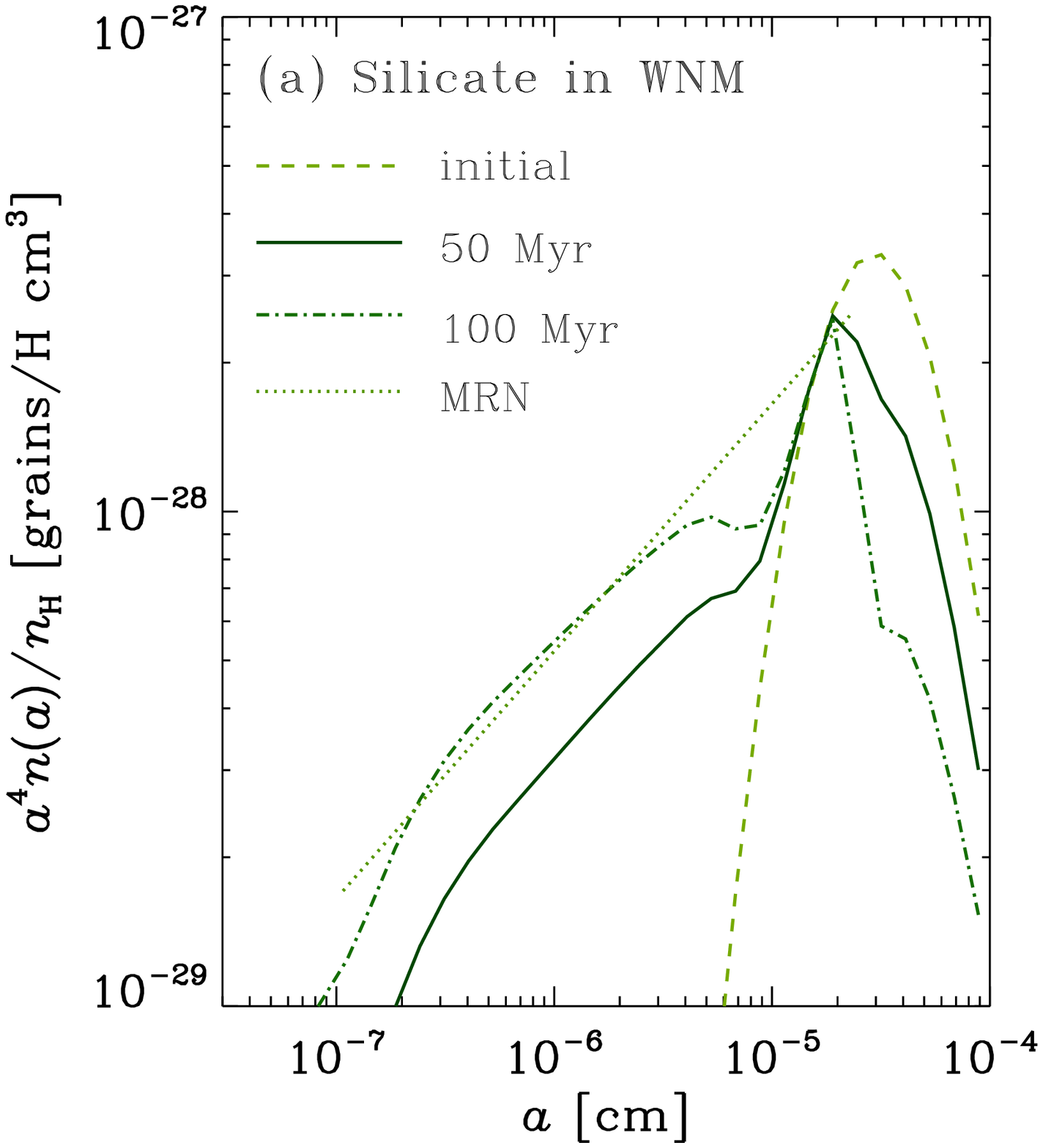}
\includegraphics[width=0.45\textwidth]{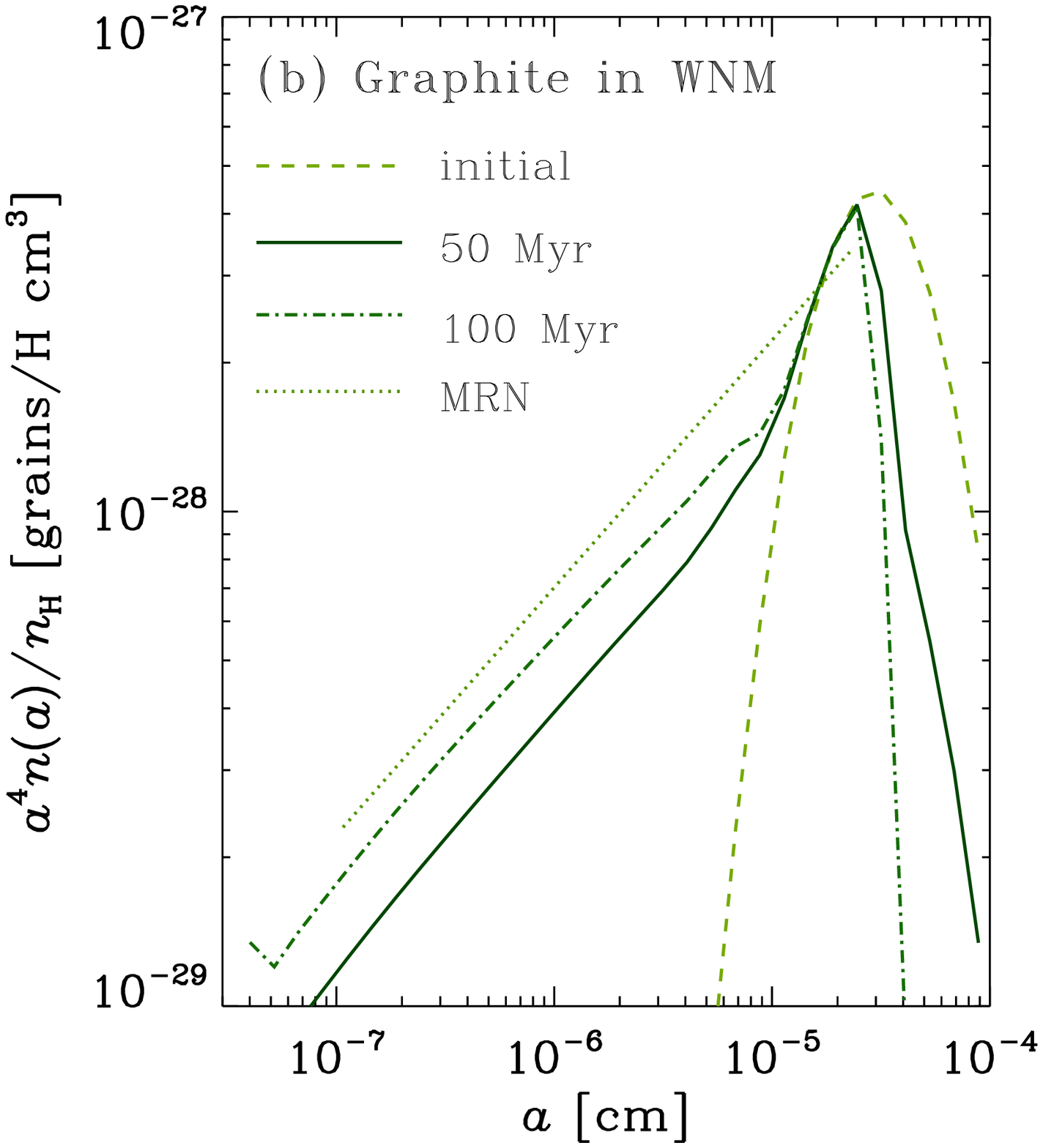}
 \caption{Grain size distributions in WNM. The grain size
 distribution is divided by $\nH$ and multiplied by
$a^4$ to show the grain mass distribution per $\log a$
and per hydrogen nucleus.
Two grain species, (a) silicate and (b) graphite,
are shown. The solid and dot-dashed lines indicate the
grain size distribution at $t=50$ and 100 Myr, respectively,
and the dashed line shows the initial log-normal
distribution. The dotted line shows the MRN size distribution.}
 \label{fig:wnm}
\end{figure*}

The evolution of grain size distribution by shattering
in WNM is shown in Fig.\ \ref{fig:wnm}.
As a reference, the MRN size distribution
($n(a)\propto a^{-3.5}$ with
$a_\mathrm{min}=10^{-3}~\micron$ and
$a_\mathrm{max}=0.25~\micron$) is also shown. We observe
that shattering redistributes the
large grains into small sizes on a time-scale of
$\la 100$ Myr,
{which is shorter than the grain lifetime in 
the ISM
(a few $\times 10^8$ yr; \citealt{jones96})}. Moreover, the
grain size distribution approaches the MRN size
distribution, which indicates
that shattering is efficient enough to enrich the small
grains up to the abundance level in the Milky Way.

Since the time-scale of shattering is inversely
proportional to the gas density and the dust-to-gas
ratio, we can summarize our result by the following
estimate of a time-scale, $\tau_\mathrm{shat}$, on
which shattering in turbulence supplies small
grains to a level compatible with the MRN size
distribution:
\begin{eqnarray}
\tau_\mathrm{shat} & \sim & 10^8(
\mathcal{R}/\mathcal{R}_0)^{-1}
(\nH /0.3\,\mathrm{cm}^{-3})^{-1}\nonumber\\
& & \times (a_0/0.1\,\micron)
(V_\mathrm{shat}/20~\mathrm{km~s}^{-1})^{-\alpha}\,
\mathrm{yr},\label{eq:taushat}
\end{eqnarray}
where $\mathcal{R}_0$ is the dust-to-hydrogen mass
ratio assumed in this paper for the Galactic
environment, and $V_\mathrm{shat}$ is the typical
relative velocity among the grains. The dependence
on $a_0$ comes from the grain collision time-scale
with a fixed total grain mass.
For the grain velocity,
$V_\mathrm{shat}\sim V_\mathrm{A}$ can be achieved
after gyroresonance. The exponent $\alpha$ is 1 if
small grain production
is only governed by the collision rate. We test
several cases where we have artificially changed the
velocities achieved by large grains, and
$\alpha$ proves to be $\sim 2$ around
$V_\mathrm{shat}\sim 5$--30 km s$^{-1}$. As mentioned in
Section \ref{subsec:coexist}, 30 per
cent of the small grains are produced by the collision
between small and large grains; in this case,
cratering occurs in large grains, and $\alpha$ becomes
larger than 1 because of the velocity dependence of
the cratered volume.

Shattering of large grains
in cold neutral medium (CNM) may also be important
if a significant fraction of the ISM is in a cold
phase \citep{mckee89}. A lower velocity
($\sim 1$--2 km s$^{-1}$; \citealt{yan04}) achieved in
CNM is compensated by a
large density ($\sim 30$ cm$^{-3}$), so that the
shattering time-scale in CNM is similar to that in WNM
(HY09) for graphite. Because of a larger shattering
threshold velocity of silicate, silicate grains
may not be shattered in CNM.

\subsection{Shattering in WIM}

\begin{figure*}
\includegraphics[width=0.45\textwidth]{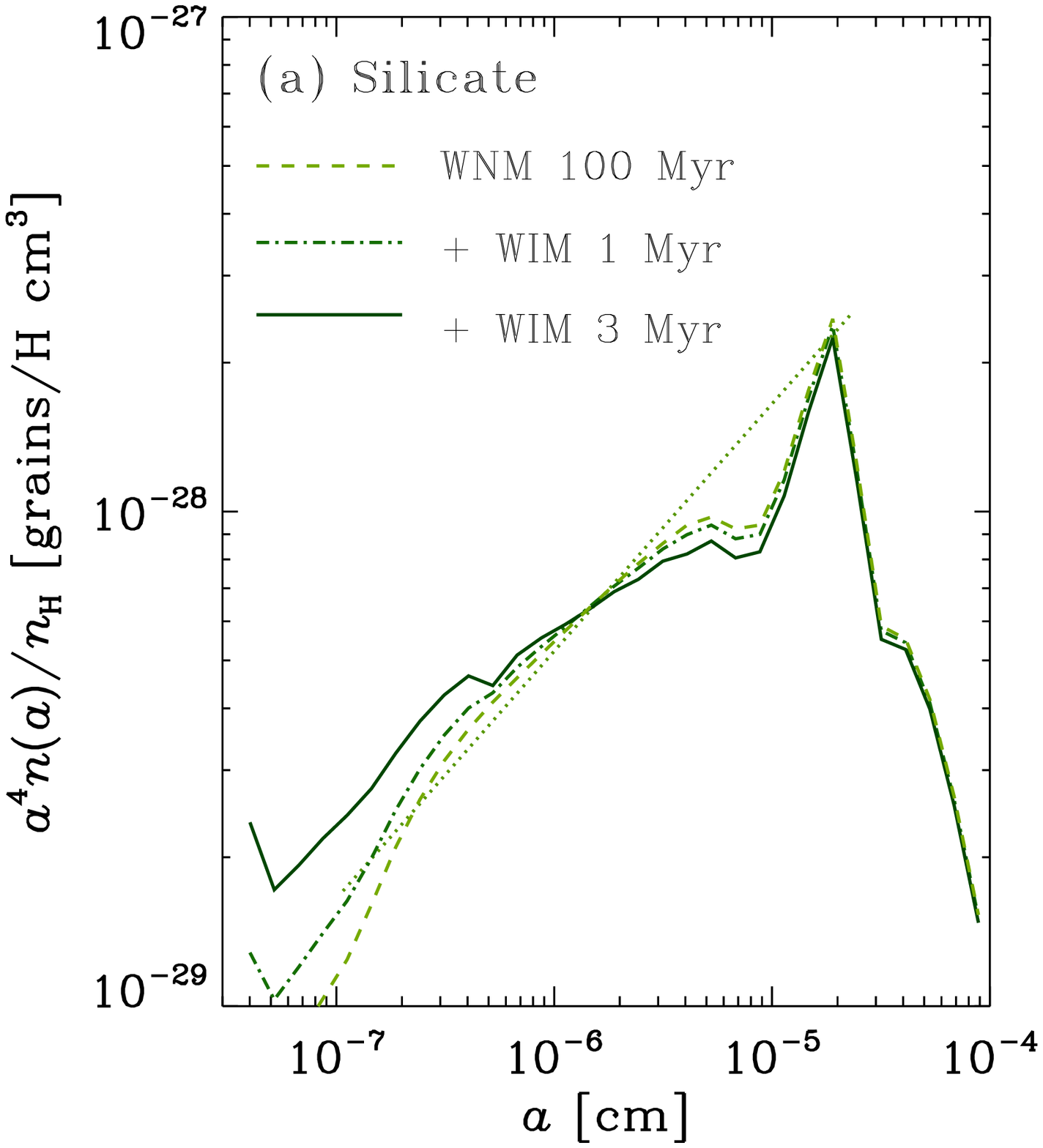}
\includegraphics[width=0.45\textwidth]{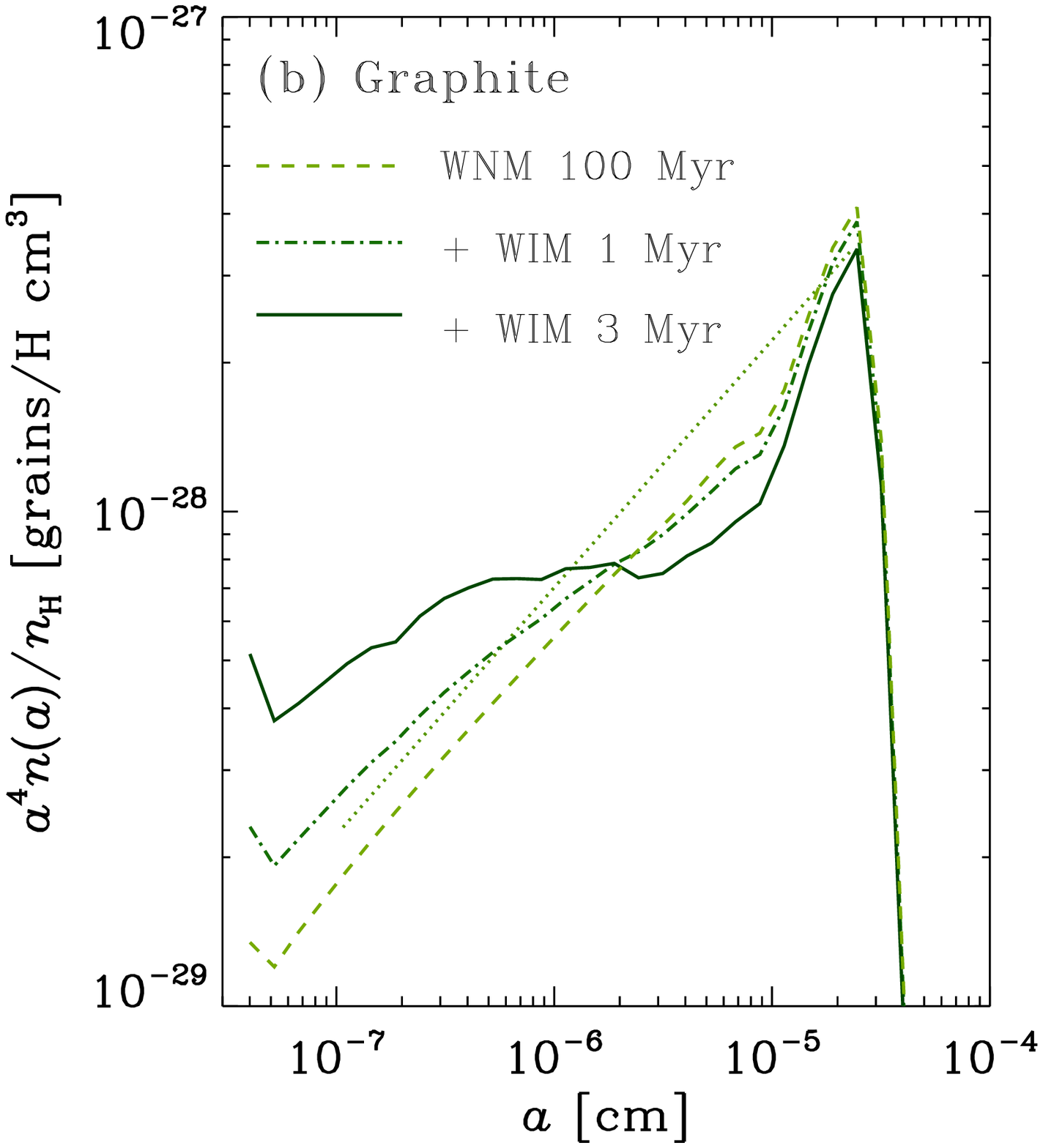}
 \caption{Same as Fig.\ \ref{fig:wnm} but with additional
 shattering in WIM. The shattering durations in WIM are 1
 and 3 Myr for the dot-dashed and solid lines,
 respectively. The dashed line shows the initial (i.e.\
 processed by shattering in WNM for 100 Myr;
 same as the dot-dashed lines in Fig.\ \ref{fig:wnm}), and
 the dotted line presents the MRN size distribution.}
 \label{fig:wim}
\end{figure*}

In this Letter, we treat shattering in WIM as an
additional effect, since larger fraction of the
volume is occupied by neutral medium. We adopt the
results at 100 Myr in Section \ref{subsec:wnm} for
the initial condition.
In Fig.~\ref{fig:wim}, we show the grain size
distribution after 1 and 3 Myr of shattering in
WIM. By shattering in WIM,
not only large grains but also relatively small grains
with $a\sim 0.01$--0.1~$\micron$ are destroyed
efficiently.
Thus, shattering in WIM may be an
effective mechanism to enhance the abundance of
small grains at $a\la 0.01~\micron$.
This is further discussed in
Section \ref{subsec:mw} in terms of the PAH
abundance.

\section{DISCUSSION}\label{sec:discussion}

\subsection{Comparison with the Galactic grain
size distributions}\label{subsec:mw}

This paper provides an experimental investigation as to
whether the mean abundance of small grains derived from
the mean extinction by MRN can be achieved by shattering.
Observationally, the grain size distribution is derived
from the extinction curve along a line of sight, which
could cross not only diffuse ISM and but also clouds.
The typical time-scale of the phase exchange between
diffuse ISM and clouds is less than 100 Myr
\citep{ikeuchi88}, shorter than the grain lifetime.
Therefore, the grains processed in WNM are mixed among
the ISM phases.
If this is true, we can directly compare the results
obtained in this paper with the grain size distribution
derived observationally from the
extinction curve.
If we consider the existence of the ISM phases other
than WNM, the shattering time-scale should be modified
as $\tau_\mathrm{shat}/f_\mathrm{WNM}$,
where $f_\mathrm{WNM}$ is the fraction of time spent in
WNM. We assume that $f_\mathrm{WNM}$ is not much smaller
than 1.

{
As shown in the previous section, shattering in WNM
redistributes large grains into small grains in the
grain lifetime so that the
abundance of small grains becomes consistent with the
MRN grain size distribution.
Therefore, shattering in turbulent ISM is a strong
candidate for the dominant production  mechanism
of small grains. The short shattering time-scale
($\tau_\mathrm{shat}\sim 100$ Myr) indicates that
the loss of large grains by shattering is
likely to be balanced with the dust growth in
dense clouds, not by the injection of large grains
from stars
(see \citealt{draine09} for the time-scales of
these processes).
}

Shattering in WIM destroys not only large grains but
also relatively small grains very efficiently. Thus,
some imprints of WIM on
the grain size distribution may be seen for the Galactic
grain size distribution. In Fig.\ \ref{fig:wim}, we
indeed observe the enhancement of small grains at
$a<100$~\AA\ for graphite. The
enhancement of very small grains is relatively
minor for silicate, since silicate is
harder to shatter than graphite. It is interesting
to point out that the enhancement of very small
grain abundance is more pronounced for
carbonaceous species than silicate species
according to
\citet{weingartner01} and \citet{li01}.
\citet{li01} have shown that
{an excess in the PAH abundance at
$a<100$ \AA\ relative to the extension of the
MRN size distribution is required
to reproduce
the PAH emission strength in the Milky Way.
Such an excess also keeps the consistency with
the extinction curve \citep{weingartner01}.}
Strictly speaking, small PAHs
should be treated as molecules as done by
\citet*{micelotta10}. Nevertheless, we expect that
PAHs once produced are not much affected by shattering
in turbulence, since small grains (also PAHs) are
coupled with small scale (i.e.\ low velocity)
turbulence.


\subsection{Coexistence of large and small grains}
\label{subsec:coexist}

Our hypothesis investigated in this paper, that is,
that shattering governs the production of very small
grains, naturally explains the ubiquitous coexistence
of large and small grains. Indeed, in the Galactic
environment, the spatial distribution of very small
grains radiating in the MIR and that of large grains
emitting in the far infrared (FIR) shows a tight
correlation \citep{shibai99}. Also the
PAH emission is always
associated with the dust continuum
\citep{onaka99}. Moreover,
the FIR colours of extragalactic objects are
consistent with the Galactic FIR colours
\citep{hibi06},
which implies that the grain size distribution is not very
different. The fact that observational MIR to
FIR SEDs can
be described by a small numbers of parameters
\citep{dale01,nagata02}
also implies that there is a ubiquitous mechanism
for determining the grain size distribution.
Finally, shattering occurs predominantly in
diffuse neutral and ionized medium, which is
consistent with the diffuse nature of the 70 $\micron$
excess coming from very small grains
\citep{bernard08}.

We have shown that even if there are only large grains,
the supply of small grains are efficient enough.
In other words, shattering by interstellar turbulence
always provides a path for small grain production.
After an enough number of small grains are produced,
it would be more
reasonable to consider that the grains suppled from
stars or dense clouds are well mixed with the MRN-like
size distribution. The shattering under an MRN size
distribution is considered in HY09, who
also obtain a similar time-scale for shattering.
{
The
similarity in the time-scale between our case and HY09's
is explained as follows: In WNM, only
large grains acquire velocities large enough for shattering.
Thus, a collision with a large grain is the only way for
a grain to be shattered.
Let us consider a `projectile' with size $a$ colliding
with a large grain with size
$a_\mathrm{large}$. The contribution from projectiles with sizes
between $a$ and $a+\mathrm{d}a$ to shattering
of the large grain is roughly proportional to
$\pi a_\mathrm{large}^2V_\mathrm{shat}n(a)\mathcal{V}\,
\mathrm{d}a$,
where $\mathcal{V}$ is the shattered volume.
If the projectile is large, $\mathcal{V}$ is large
(in particular, if
$a\sim 0.1~\micron$, $\mathcal{V}$ is equal
to the
entire volume of the large grain). Thus, larger projectiles
have larger impacts on the large grain.
Indeed, if we consider shattering among large
($a>0.08~\micron$) grains only, the production rate
of grains with $a\sim 0.01~\micron$ becomes only
30 per cent less (i.e.\ 30 per cent of small grain production
is attributed to the collisions between small and
large grains).
Since $n(a)$ is similar between our initial distribution
and the MRN at large sizes, we obtain a similar
shattering time-scale to HY09.
}

\subsection{Metallicity dependence}

Since the shattering time-scale is inversely
proportional to the dust-to-gas ratio
(equation \ref{eq:taushat}), it is expected that the
abundance of very small grains is low in dust-poor
(metal-poor) galaxies.
However, as pointed out by \citet{hi09}, detection of
infrared dust emission from low-metallicity objects
could be biased to dense systems. In such dense
environments, shattering is enhanced.
Consequently, the MIR radiation from very small grains
may become strong \citep{galliano05,engelbracht08}. Thus,
dust emission from diffuse (low-density)
low-metallicity
system can be a strong test for our theory:
if the MIR continuum is suppressed in
such a system, our hypothesis that shattering is
the production source of very small grains is
supported.

The production of small grains by shattering also
indicates that the materials of small grains reflect
those of original large grains. The formation
mechanism of large grains should depend on
metallicity: The production
of large grains by accretion of heavy elements
in molecular
clouds is inefficient in low metallicity. Thus, stellar
production of grains should be the dominant
source of large grains in low-metallicity
environments \citep[e.g.][]{matsuura08,galliano08}.
Considering that molecular clouds
are the site of various chemical reactions,
large grains formed by stars and those grown in
molecular clouds could be chemically different.
The metallicity dependence of MIR
emission properties, especially that of PAH emission
\citep[e.g.][]{smith07,draine07},
may reflect the different properties of original large
grains in terms of metallicity. Also if there is a spatial
variation of production sources of large grains,
spatial variation of very small grains and PAHs
may also be present as observed by
\citet{bernard08} and \citet{paradis09}.

\section{Conclusion}\label{sec:conclusion}

We have theoretically investigated interstellar
shattering of large grains ($a\sim 0.1~\micron$) as
a production source of small grains. We have shown
that by shattering in WNM the abundance of small
grains reaches the level consistent with the MRN
size distribution in $\sim 10^8$ yr (i.e.\ in
the grain lifetime). Shattering in WIM additionally
destroys grains with $a\sim 0.01~\micron$ and
redistribute them into
smaller sizes. This effect is more pronounced for the
carbonaceous grains, and can explain the
enhancement of very small carbonaceous grains
(or PAHs) with $a<100$\,\AA\ as indicated observationally
by \citet{li01}. Since turbulence is ubiquitous
in the ISM, our theory naturally explains the
strong MIR--FIR correlation observed generally
for galaxies.

\section*{Acknowledgments}
We thank T. Nozawa and T. Kozasa for helpful
discussions, and H. Yan
for providing the data of grain velocities in turbulence.



\bsp

\label{lastpage}


\begin{thebibliography}{99}
\bibitem[\protect\citeauthoryear{Bernard et al.}{2008}]{bernard08}
    Bernard, J.-P., et al.\ 2008, AJ, 136, 919
\bibitem[\protect\citeauthoryear{Bianchi \& Schneider}{2007}]{bianchi07}
    Bianchi, S., \& Schneider, R. 2007, MNRAS, 378, 973
\bibitem[\protect\citeauthoryear{Borkowski \& Dwek}{1995}]{borkowski95}
    Borkowski, K. J., \& Dwek, E. 1995, ApJ, 454, 254
\bibitem[\protect\citeauthoryear{Dale et al.}{2001}]{dale01}
    Dale, D. A., Helou, G., Contursi, A., Silbermann, N. A., \& Kolhatkar, S.
    2001, ApJ, 549, 215
\bibitem[\protect\citeauthoryear{Draine}{2009}]{draine09}
    Draine, B. T. 2009, in Henning Th., Gr\"{u}n E., Steinacker J., eds.,
    ASP Conf.\ Ser., Vol.\ 414, Cosmic Dust --- Near and Far, ASP,
    San Francisco, 453
\bibitem[\protect\citeauthoryear{Draine et al.}{2007}]{draine07}
    Draine, B. T., et al.\ 2007, ApJ, 663, 866
\bibitem[\protect\citeauthoryear{Dwek}{1998}]{dwek98}
    Dwek, E. 1998, ApJ, 501, 643
\bibitem[\protect\citeauthoryear{Engelbracht et al.}{2008}]{engelbracht08}
    Engelbracht, C. W., Rieke, G. H., Gordon, K. D., Smith, J.-D. T.,
    Werner, M. W., Moustakas, J., Willmer, C. N. A., \& Vanzi, L. 2008,
    ApJ, 678, 804
\bibitem[\protect\citeauthoryear{Gail, Keller, \& Sedlmayr}{1984}]{gail84}
    Gail, H.-P., Keller, R., \& Sedlmayr, E. 1984, A\&A, 133, 320
\bibitem[\protect\citeauthoryear{Galliano, Dwek, \& Chanial}{2008}]{galliano08}
    Galliano, F., Dwek, E., \& Chanial, P. 2008, ApJ, 672, 214
\bibitem[\protect\citeauthoryear{Galliano et al.}{2005}]{galliano05}
    Galliano, F., Madden, S. C., Jones, A. P., Wilson, C. D., \&
    Bernard, J.-P. 2005, A\&A, 434, 867
\bibitem[\protect\citeauthoryear{Gauger et al.}{1999}]{gauger99}
    Gauger, A., Balega, Y. Y., Irrgang, P., Osterbart, R., \& Weigelt, G. 1999,
    A\&A, 346, 505
\bibitem[\protect\citeauthoryear{Groenewegen}{1997}]{groenewegen97}
    Groenewegen, M. A. T. 1997, A\&A, 317, 503
\bibitem[\protect\citeauthoryear{Hibi et al.}{2006}]{hibi06}
    Hibi, Y., Shibai, H., Kawada, M., Ootsubo, T., \& Hirashita, H.
    2006, PASJ, 58, 509
\bibitem[\protect\citeauthoryear{Hirashita et al.}{2010}]{hirashita10}
    Hirashita, H., Nozawa, T., Yan, H., \& Kozasa, T. 2010,
    MNRAS, in press (arXiv:1001.2606)
\bibitem[\protect\citeauthoryear{Hirashita \& Ichikawa}{2009}]{hi09}
    Hirashita, H., \& Ichikawa, T. T. 2009, MNRAS, 396, 500
\bibitem[\protect\citeauthoryear{Hirashita \& Yan}{2009}]{hirashita09}
    Hirashita, H., \& Yan, H. 2009, MNRAS, 394, 1061 (HY09)
\bibitem[\protect\citeauthoryear{Hofmann et al.}{2001}]{hofmann01}
    Hofmann, K.-H., Balega, Y., Bl\"{o}cker, T., \& Weigelt, G. 2001, A\&A,
    379, 529
\bibitem[\protect\citeauthoryear{Ikeuchi}{1988}]{ikeuchi88}
    Ikeuchi, S. 1988, Fundam. Cosm. Phys., 12, 255
\bibitem[\protect\citeauthoryear{Jones, Tielens, \& Hollenbach}{Jones et al.}{1996}]{jones96}
    Jones, A. P., Tielens, A. G. G. M., \& Hollenbach, D. J. 1996,
    ApJ, 469, 740
\bibitem[\protect\citeauthoryear{Jones et al.}{1994}]{jones94}
    Jones, A. P., Tielens, A. G. G. M., Hollenbach, D. J., \& McKee, C. F.
    1994, ApJ, 433, 797
\bibitem[\protect\citeauthoryear{Kozasa, Hasegawa, \& Nomoto}{1989}]{kozasa89}
    Kozasa, T., Hasegawa, H., \& Nomoto, K. 1989, ApJ, 344, 325
\bibitem[\protect\citeauthoryear{Li \& Draine}{2001}]{li01}
    Li, A., \& Draine, B. T. 2001, ApJ, 554, 778
\bibitem[\protect\citeauthoryear{Mathis, Rumpl, \& Nordsieck}{1977}]{mathis77}
    Mathis, J. S., Rumpl, W., \& Nordsieck, K. H. 1977, ApJ, 217, 425
    (MRN)
\bibitem[\protect\citeauthoryear{Matsuura et al.}{2008}]{matsuura08}
    Matsuura, M., et al.\ 2008, in Kwok S., Sandford S., eds., IAU Symp.\ 251,
    Organic Matter in Space, Cambridge Univ.\ Press, Cambridge, p.\ 197
\bibitem[\protect\citeauthoryear{McKee}{1989}]{mckee89}
    McKee, C. F. 1989, in Allamandola L. J., Tielens, A. G. G. M., eds.,
    IAU Symp.\ 135, Interstellar dust, Kluwer, Dordrecht, p.\ 431
\bibitem[\protect\citeauthoryear{Micelotta, Jones, \& Tielens}{Micelotta et al.}{2010}]{micelotta10}
    Micelotta, E. R., Jones, A. P., Tielens, A. G. G. M. 2010, A\&A,
    510, 36
\bibitem[\protect\citeauthoryear{Nagata et al.}{2002}]{nagata02}
    Nagata, H., Shibai, H., Takeuchi, T. T., \& Onaka, T. 2002, PASJ,
    54, 695
\bibitem[\protect\citeauthoryear{Nozawa et al.}{2007}]{nozawa07}
    Nozawa, T., Kozasa, T., Habe, A., Dwek, E., Umeda, H.,
    Tominaga, N., Maeda, K., \& Nomoto, K. 2007, ApJ, 666, 955
\bibitem[\protect\citeauthoryear{O'Donnell \& Mathis}{1997}]{odonnell97}
    O'Donnell, E., \& Mathis, J. S. 1997, ApJ, 479, 806
\bibitem[\protect\citeauthoryear{Onaka et al.}{1999}]{onaka99}
    Onaka, T.,  Mizutani, M., Tomono, D., Shibai, H., Nakagawa, T., \&
    Doi, Y. 1999, in Cox P., Kessler M. F. eds., ESA-SP 427, The
    Universe as Seen by ISO, ESA, Paris, p.\ 731
\bibitem[\protect\citeauthoryear{Paradis et al.}{2009}]{paradis09}
    Paradis, D., et al.\ 2009, AJ, 138, 196
\bibitem[\protect\citeauthoryear{Shibai, Okumura, \& Onaka}{1999}]{shibai99}
    Shibai, H., Okumura, K., \& Onaka, T. 1999, in Nakamoto T. ed.,
    Star Formation 1999, Nobeyama Radio Observatory, Nobeyama, p.\ 67
\bibitem[\protect\citeauthoryear{Smith et al.}{2007}]{smith07}
    Smith, J. D. T., et al.\ 2007, ApJ, 656, 770
\bibitem[\protect\citeauthoryear{Weingartner \& Draine}{2001}]{weingartner01}
    Weingartner, J. C., \& Draine, B. T. 2001, ApJ, 548, 296
\bibitem[\protect\citeauthoryear{Yan \& Lazarian}{2003}]{yan03}
    Yan, H., \& Lazarian, A. 2003, ApJ, 592, L33
\bibitem[\protect\citeauthoryear{Yan, Lazarian, \& Draine}{Yan et al.}{2004}]{yan04}
    Yan, H., Lazarian, A., \& Draine, B. T. 2004, ApJ, 616, 895
\end{thebibliography}
\end{document}